%% file: main.tex
\documentclass[sigconf]{acmart}

\input{misc/customize.tex}

\AtBeginDocument{%
  }

\copyrightyear{2026}
\acmYear{2026}
\setcopyright{cc}
\setcctype{by}
\acmConference[ISPD '26]{Proceedings of the 2026 International Symposium on Physical Design}{March 15--18, 2026}{Bonn, Germany}
\acmBooktitle{Proceedings of the 2026 International Symposium on Physical Design (ISPD '26), March 15--18, 2026, Bonn, Germany}
\acmDOI{10.1145/3764386.3780044}
\acmISBN{979-8-4007-2314-8/2026/03}

\begin{document}

\title[Toward Accurate, Large-scale Electromigration Analysis and Optimization in Integrated Systems]{Invited: Toward Accurate, Large-scale Electromigration\\
Analysis and Optimization in Integrated Systems}

\author{Sachin S. Sapatnekar}
\affiliation{%
  \institution{University of Minnesota}
  \city{Minneapolis}
  \state{MN}
  \country{USA}
}


\begin{abstract}
Electromigration, a significant lifetime reliability concern in high-performance integrated circuits, is projected to grow even more important in future heterogeneously integrated systems that will service higher current loads. Today, EM checks are primarily based on rule-based methods, but these have known limitations. In recent years, there has been remarkable progress in enabling fast EM computations based on more accurate physics-based models, but such methods have not yet moved from research to practice. This paper overviews physics-based EM models, contrasts them with empirical models, and outlines several open problems that must be solved in order to enable accurate physics-based and circuit-aware EM analysis and optimization in future integrated systems.
\end{abstract}




\maketitle

\input{sec/1-Intro}
\input{sec/2-Traditional}
\input{sec/3-Physics}
\input{sec/4-Open-Analysis}
\input{sec/5-Optimization}
\input{sec/6-Conclusion}

\section*{Acknowledgments}

The author gratefully acknowledges numerous learnings from past and current collaborators: (alphabetically) Nestor Evmorfopoulos, Palkesh Jain, Vivek Mishra, Gracieli Posser, Susann Rothe, and Mohammad Shohel. This work is supported in part by the National Science Foundation and by the SRC JUMP 2.0 CHIMES Center.

\bibliographystyle{ACM-Reference-Format}
\bibliography{ref}

\end{document}

%% file: misc/customize.tex
\usepackage{amsmath}
\usepackage{enumitem}
\usepackage{hhline}
\usepackage{graphicx}
\usepackage{multirow}
\usepackage{color}
\usepackage{fancybox}
\usepackage{url}
\usepackage[normalem]{ulem}
\usepackage{siunitx}
\usepackage{caption}
\usepackage{algorithm}
\usepackage{subfig}

\definecolor{gray1}{gray}{0.7}
\definecolor{gray2}{gray}{0.98}
\definecolor{light-gray}{gray}{0.95}

\newcommand{\ignore}[1]{}

\newcommand{\Zeff}{Z^\star_{\mbox{\scriptsize \em eff}}}
\newcommand{\jLcrit}{(jL)_{\mbox{\scriptsize \em crit}}}
\newcommand{\Javg}{J_{\mbox{\scriptsize \em avg}}}
\newcommand{\tlife}{t_{\mbox{\scriptsize \em life}}}

%% file: sec/1-Intro.tex
\section{Introduction}
\label{sec:introduction}

\noindent
Electromigration (EM)~\cite{Lienig25} has long been recognized as a significant source of interconnect failures in integrated circuit design. The phenomenon is caused by the flow of high currents through wires over long periods, which causes a transfer of momentum from charge carriers to metal atoms. Eventually, this causes atoms to migrate, and results in the creation of voids at the cathode and/or hillocks at the anode.  Studies of the problem date back to at least six decades~\cite{Grone61, Penney64}, and the problem was actively researched in the context of integrated circuits (ICs) during the period where the dominant interconnect metal was aluminum~\cite{black:69b}.  Through the transition from aluminum to copper, the issue has persisted, particularly for narrower wires in lower metal layers~\cite{ala:05,Auth17}. However, unlike aluminum interconnects, atoms in a copper metallization technology are prevented from migrating across metal layers due to the presence of barrier material in dual-damascene copper processes. 

\begin{figure}[t]
\centering
\subfloat[]{
\includegraphics[width=0.57\linewidth]{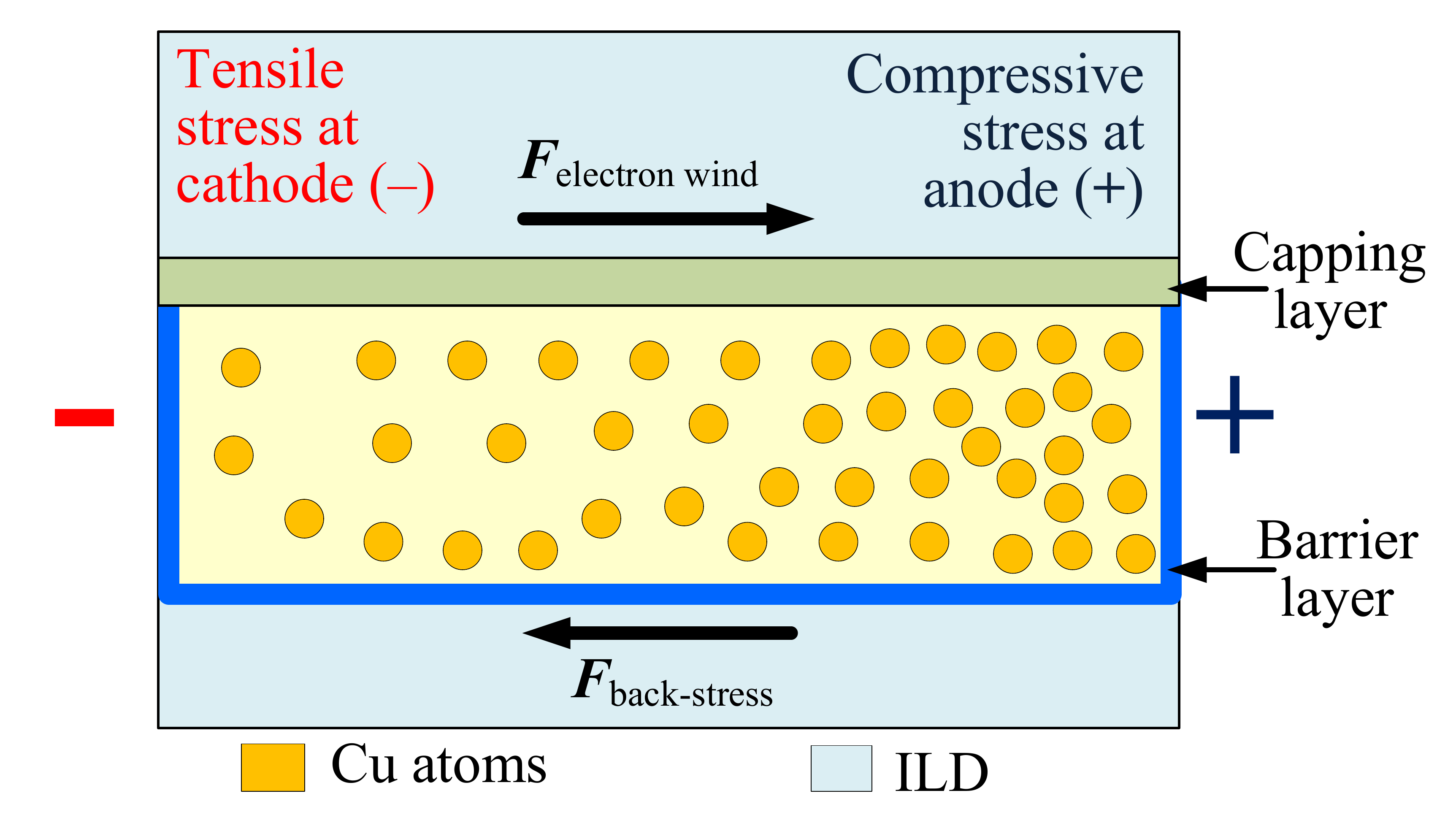}
}
\hspace{0.03\linewidth}
\subfloat[]{
\includegraphics[width=0.35\linewidth]{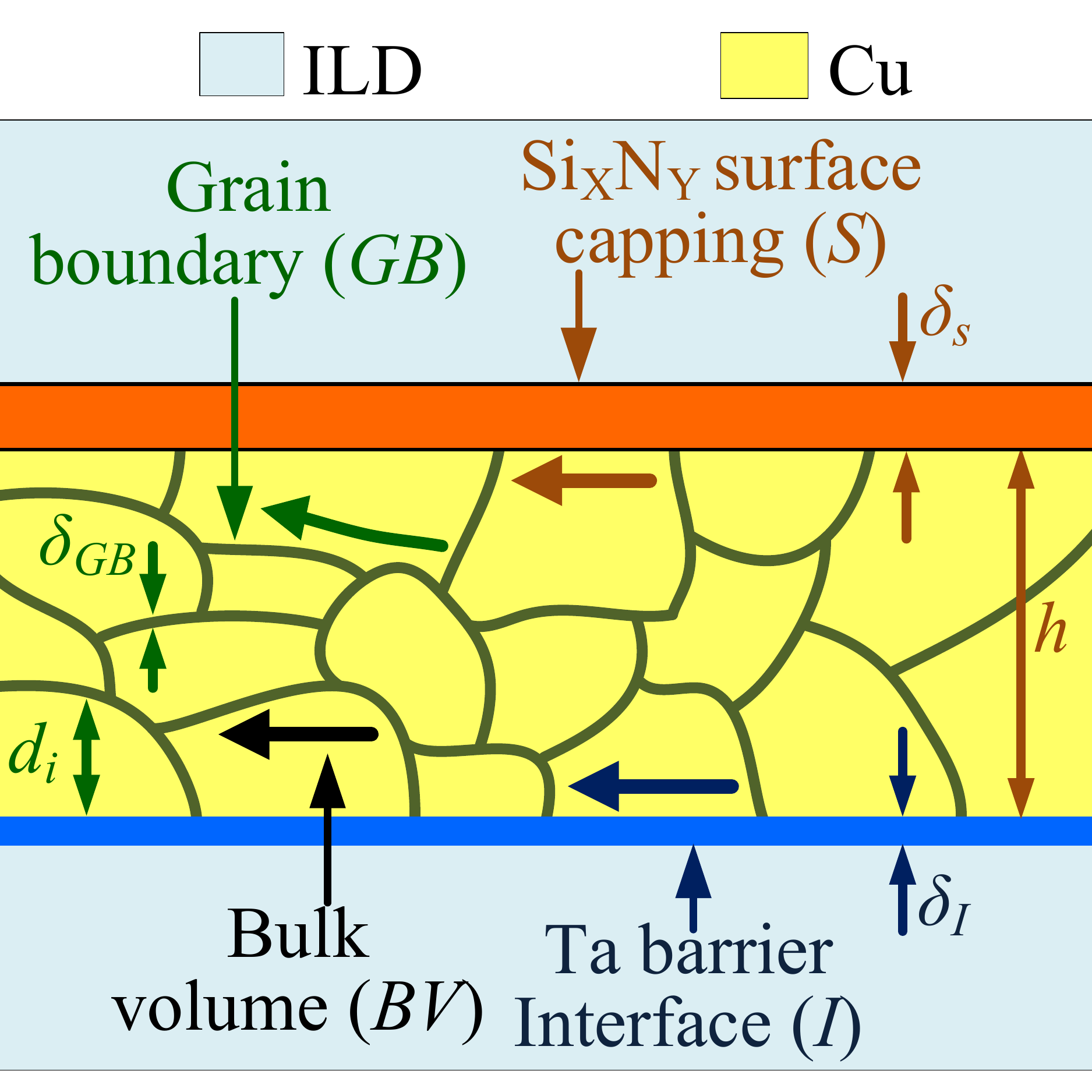}
}
\vspace{-4mm}
\caption{(a)~Cross section of a Cu wire indicating the back-stress and the electron wind force~\cite{vivek:dac}. (b)~Atomic diffusion paths in a copper interconnect.}
\Description{Cross section of a Cu wire indicating the back-stress and the electron wind force~\cite{vivek:dac}.}
\vspace{-4mm}
\label{fig:schem}
\end{figure}

Today, metal migration in ICs is understood to be caused by three primary factors, illustrated for simplicity in Fig.~\ref{fig:schem} for a single-segment interconnect, with electron current (opposite to the direction of conventional current) flowing from a cathode to an anode:
\begin{itemize}
\item The \textit{electron wind}, associated with the current density in a wire, creates a force that directs metal atoms from the cathode towards the anode.
\item The \textit{back-stress} creates an opposing force, that increases as the concentration gradient of atoms grows nonuniform, with more atoms towards the anode than the cathode.
\item The \textit{residual stress} in a wire is caused as the interconnect stack is cooled from a higher manufacturing temperature to room temperature or chip operating temperatures, and can be attributed to differences in the coefficients of thermal expansion (CTEs) of materials in the interconnect stack.
\end{itemize}
The interactions of these three factors creates a stress distribution within the wire that can be described by a partial differential equation (PDE), with the residual stress acting as an initial condition for the PDE. The movement of migrated atoms is blocked at either end of the wire, and at vias, due to the atom-impermeable barrier layer. The electron wind force causes atomic depletion at the cathode, resulting in the generation of tensile stress at the cathode; at the anode, the migrated atoms accumulate, creating compressive stress, as shown in Fig.~\ref{fig:schem}(a).  The microstructure of the metallization consists of grains of metal, and atoms diffuse through four mechanisms, illustrated in Fig.~\ref{fig:schem}(b): along the grain boundaries (GB), the capping surface (S), the bulk volume (BV), and the interface barrier (I), with the former two mechanisms dominating the others due to their lower activation energies~\cite{hu:99}.

A small stress results in proportional strain (with Young's modulus being the constant of proportionality) that does not significantly change wire resistance, but a large stress, exceeding the elastic limit, can nucleate a void in the wire. Under these circumstances, the wire resistance rises rapidly. In copper interconnects, with cladding and barrier layers, a very small void still has a conducting path through these layers (but nevertheless results in a resistance increase), but as the void grows, the resistance rises to be unacceptably high, effectively signaling a break in the wire. Typically, when individual interconnects are analyzed, an EM-induced resistance increase of 10\% is considered as the threshold for the failure of a wire. However, as will be pointed out later, various redundant systems, including power grids and clock meshes, can withstand larger degradations and continue to function correctly.

Beyond digital circuits, EM is increasingly seen in a number of applications today and promises to grow further in the future. Analog circuits often carry DC bias currents for long periods of time, and these can result in electromigration.  Reliability verification and EM fixes are an important part of modern analog design methodologies.  A second emerging application is in advanced packaging with three-dimensional (3D) circuits and heterogeneous integration~(HI), placing multiple chiplets on a substrate and connecting them with dense interconnects (e.g., EMIB, CoWoS, Foveros) over one of multiple substrate types (organic, silicon, glass). HI provides a viable alternative for integrating large systems at low cost and high yield~\cite{HIR24}.  EDA challenges in HI systems~\cite{Sapatnekar25} include reliability factors related to EM in interconnects within the HI package, e.g., in wires, through-silicon vias (TSVs), and microbumps; EM in HI is exacerbated by high temperatures in the package, as HI systems can experience high power dissipation.

This paper summarizes a set of recent accomplishments in advancing physics-based methods, including results that have delivered efficient, linear-time solutions for both the steady-state and transient EM analysis problem in interconnects. These methods are not only well-grounded theoretically but also allow interpretability of the results, unlike opaque empirical methods.  Section~\ref{sec:acdc} begins by describing AC and DC EM, followed by a description of traditional and physics-based analyzers in Sections~\ref{sec:traditional} and \ref{sec:physics}, respectively, for individual interconnects.\footnote{Machine learning models for EM analysis have also been proposed in various contexts, including solving individual interconnect trees, for stochastic analysis~\cite{Lamichhane23}, and for identifying EM hotspots~\cite{todaes-chhabria21,Hou23}, but are not described in this paper: analytical first-principles methods already provide computationally cheap evaluations.} These ideas are then extended to system-level analysis in Section~\ref{sec:circuits}.  Next, Sections~\ref{sec:open-analysis} and \ref{sec:open-optimization} describe some open problems in analysis, including calibrating physics-based models with measurement data, and optimization that can leverage recent advances, and are followed by concluding remarks in Section~\ref{sec:conclusion}. Throughout the paper, several caveats are shared, warning the reader of pitfalls that took the author many years to comprehend.

\section{DC and AC EM failures}
\label{sec:acdc}

In power grid wires, currents predominantly flow in the same direction, implying that the electron wind drives metal migration in the same direction.  For signal wires (including clock lines), the passage of current in opposite directions in a wire results in alternate periods of EM degradation and recovery. This is because the forward current flow in one direction changes the atomic distribution in the wire, and the reverse current heals some of this atomic restructuring. This is often referred to as ``AC EM'': a recovery factor $r$ models this effect~\cite{Hunter97,lee:recovery} and determines an ``effective current density'' for the signal wire, which corresponds to the equivalent DC current density for these bidirectional signals. The expressions for DC EM are then used for AC EM modeling, with the effective current density replacing the DC current density. However, it is important to emphasize that AC and DC EM are manifested in different ways. In DC EM, the current direction is unchanged, the cathode and anode locations stay the same, and void formation is seen near one end, the cathode. In contrast, for AC EM, the cathode and anode are exchanged whenever the current direction reverses. This leads to the possibility of void nucleation at either end of the copper wire segment in a given layer.

\begin{figure}[t]
\centering
\includegraphics[width=0.5\linewidth]{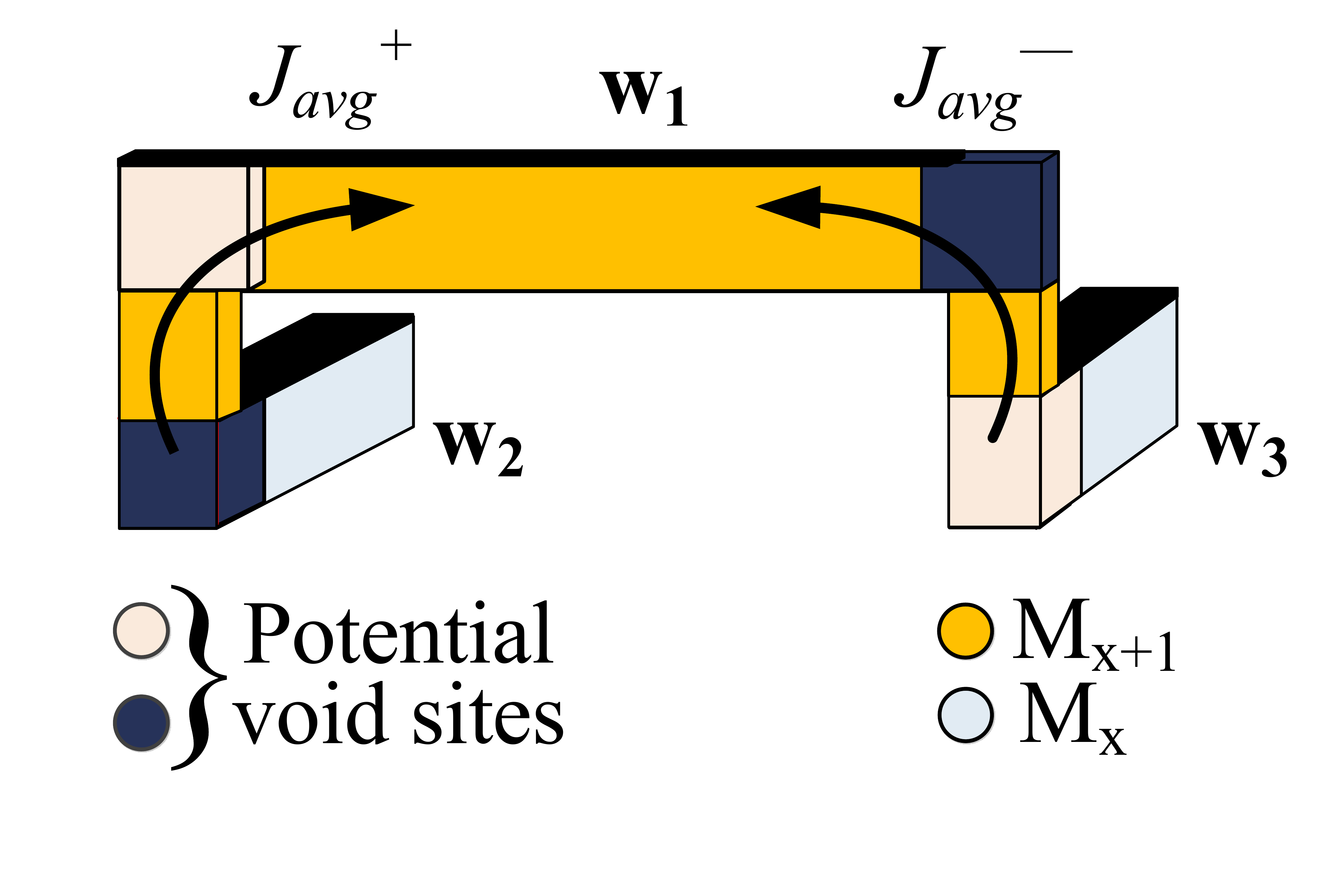}
\vspace{-6mm}
\caption{Potential void locations for AC EM.}
\Description{Potential void locations for AC EM.}
\vspace{-4mm}
\label{fig:voidsingle}
\end{figure}

For a wire carrying a bidirectional current (Figure~\ref{fig:voidsingle}), in the forward cycle, due to the electron current density $\Javg^+$, electrons flow from left to right in wire $\rm \bf w_1$ in metal layer M$_{\rm x+1}$, leading to potential void sites at the left end of $\rm \bf w_1$ and below the via in wire $\rm \bf w_3$ on layer M$_{\rm x}$. For $\rm \bf w_1$, let $J_{\mbox{\tiny EM}}^{ \mbox{\tiny L}}$ denote the effective current density at the left end of the wire.  In the reverse cycle, the current density $\Javg^-$ may cause voids at the right end of $\rm \bf w_1$ and under the via in $\rm \bf w_2$, with the effective current density at the right end of the wire being $J_{\mbox{\tiny EM}}^{\mbox{\tiny R}}$. These effective current densities are given by:
\begin{eqnarray} 
J_{\mbox{\tiny EM}}^{ \mbox{\tiny L}} = \Javg^+ - r \cdot \Javg^- \quad \text{and} \quad
J_{\mbox{\tiny EM}}^{ \mbox{\tiny R}} = \Javg^- - r \cdot \Javg^+.
\end{eqnarray} 

\ignore{
For $\rm \bf w_1$, the effective current density at the left end of the wire, $J_{\mbox{\tiny EM}^{ \mbox{\tiny L}}}$ is given by: 
\begin{eqnarray} 
J_{\mbox{\tiny EM}^{ \mbox{\tiny L}}} &=& \Javg^+ - r \cdot \Javg^-
\label{eq:ourj1} 
\end{eqnarray}
where $r$ is the recovery factor \cite{lee:recovery}. In the reverse cycle, the current density $\Javg^-$ may cause voids at the right end of $\rm \bf w_1$ and under the via in $\rm \bf w_2$. the effective current density at the right end of the wire, $J_{\mbox{\tiny EM}^{\mbox{\tiny R}}}$, is given by:
\begin{eqnarray}
J_{\mbox{\tiny EM}^{ \mbox{\tiny R}}} &=& \Javg^- - r \cdot \Javg^+
\label{eq:ourj2}
\end{eqnarray}
}

\noindent
\underline{\textit{Caveat 1}}
Due to differences in the sizes of PMOS and NMOS devices, which drive rise and fall currents, respectively, as well as differences in electron and hole mobilities, the forward and reverse current densities may be quite different. As a result, the effective current density could be nonzero even if $r=1$ (recall that $r=1$ allows for full recovery under symmetric currents).

%% file: sec/2-Traditional.tex
\section{Traditional EM Models}
\label{sec:traditional}

\subsection{Black's equation and the Blech criterion}

\noindent
Today's EM methodologies are largely based on Black's equation~\cite{black:69b}:
\begin{equation}
t_{50} = \frac{A}{j^n}\exp{\frac{E_a}{kT}}
\label{eq:black}
\end{equation}
Here, $t_{50}$ is the mean time to failure (MTTF) for the wire;\footnote{In \cite{black:69b}, this equation characterized MTF, the median time to failure, but it is more customary to work with the mean time to failure today.} $A$ and $n$ are constants; $j$ is the current density in the wire; $E_a$ is the  activation energy; $k$ is Boltzmann's constant; and $T$ is the temperature. 

In~\cite{black:69b}, the form of the equation was derived from theoretical principles.  This derivation used a value of $n=2$, but later work revised the value of $n$ to lie between 1 and 2, incorporating effects that were not considered in the original derivation~\cite{lloyd:07}. Using Black's equation, the lognormal distribution of the MTTF is semi-empirically extracted, based on measurements on test structures corresponding to single-segment wires.  The failure times of these test structures, exposed to high currents and elevated temperatures to accelerate EM-induced failure (in the order of hours), are extrapolated to normal operating conditions to obtain the time to failure (in the order of years), and are used to calibrate Black's equation. 

Failure rate requirements depend on the market (consumer, industrial, military): typical values are provided in~\cite{Jain16}.  Chip reliability engineers translate a chip-level specification to specific fail fraction (FF) targets, in units of failures-in-time (FITs), on individual wires.  The FF corresponds to a point $z$ on the underlying normal cumulative distribution function (CDF) of the lognormal:
\begin{align}
\text{FF} = \Phi (z); \quad z = \frac{\ln t_f - \ln t_{50}}{\sigma},
\label{eq:FF2z}
\end{align}
where $\Phi(.)$ is the CDF of the standard normal distribution $N(0,1)$, and $\sigma$ is the standard deviation of the underlying Gaussian.

\ignore{
The FF corresponds to a point $z$ on the CDF of a lognormal distribution, whose PDF is given by
\begin{align}
f(t) = \frac{1}{\sigma t\sqrt{2 \pi}} e^{-\frac{\ln t_f - \ln t_{50}^2}{2\sigma^2}} dx
\end{align}
where $\sigma$ is the standard deviation.  We can thus relate FF to the CDF as
\begin{align}
\text{FF} = \frac{1}{\sqrt{2 \pi}} \int_{-\infty}^z e^{-x^2/2} dx
\end{align}
where $z = \frac{\ln(t_f) - \ln(t_{50})}{\sigma}$.
}

Using~\eqref{eq:FF2z}, the market-specific fail fraction specification, FF, is translated to a value of $z$. For a given product lifetime requirement, $t_f$, and the characterized variance $\sigma$, it is possible to obtain a minimum specification on $t_{50}$.  Substituting this in~\eqref{eq:black}, we obtain the \textit{maximum allowable current density} in a wire.

Further work by Blech, also on single-segment interconnects, discovered that the interconnect length also has an impact on MTTF~\cite{blech:76}: shorter interconnects with the same current density are less likely to fail.  More specifically, the threshold current density required to induce EM varies inversely with the wire length. The derivation in~\cite{blech:76} showed that the forward electron wind force in a line segment is opposed by the back-stress described in Section~\ref{sec:introduction} due to differences in the spatial concentration of metal atoms along the line. For shorter wires, the two cancel each other out in the steady state, and the wire is ``immortal'' (i.e., immune to EM) if the maximum steady-state stress in the wire is insufficient to create a void. In longer wires and in wires with larger current densities, the electron wind force may be stronger and can lead to void formation. Specifically, the criterion for immortality is that the product of the current density and the wire length $L$ must fall below a technology-specific number denoted as $\jLcrit$:
\begin{equation}
    j L < \jLcrit 
    \label{eq:blech}
\end{equation}
This criterion is obtained by equating the electron wind force to the (growing) back stress force, corresponding to the steady-state case where the two forces cancel each other out and the maximum stress is achieved: if this stress is below the critical stress $\sigma_c$ that causes a rupture in the wire, then the wire is considered immortal. Blech's derivation shows that $\jLcrit = (2 \Omega/(e\Zeff\rho)) \sigma_c$, i.e., the limit is a function of $\sigma_c$. The other symbols in the expression for $\jLcrit$ are material or physical constants, defined in Section~\ref{sec:em_equations}.

\subsection{EM checks based on current densities}

\noindent
Although initially conceived for aluminum-based interconnects, Black's-equation-based semi-empirical methodologies continue to be used in the copper era. A typical EM check~\cite{Mishra16} may estimate current densities, use Black's equation to flag wires whose current density exceeds the specified threshold, and then use the Blech criterion~\eqref{eq:blech} to eliminate any flagged wires that are immortal.  An alternative, widely used, method checks the wire current density against a threshold regardless of its length -- but as a nod to the Blech criterion, wires below a threshold length are subject to larger current density limits.

%% file: sec/3-Physics.tex
\section{Physics-based EM models}
\label{sec:physics}

\begin{figure}[t]
\centering
\includegraphics[width=0.6\linewidth]{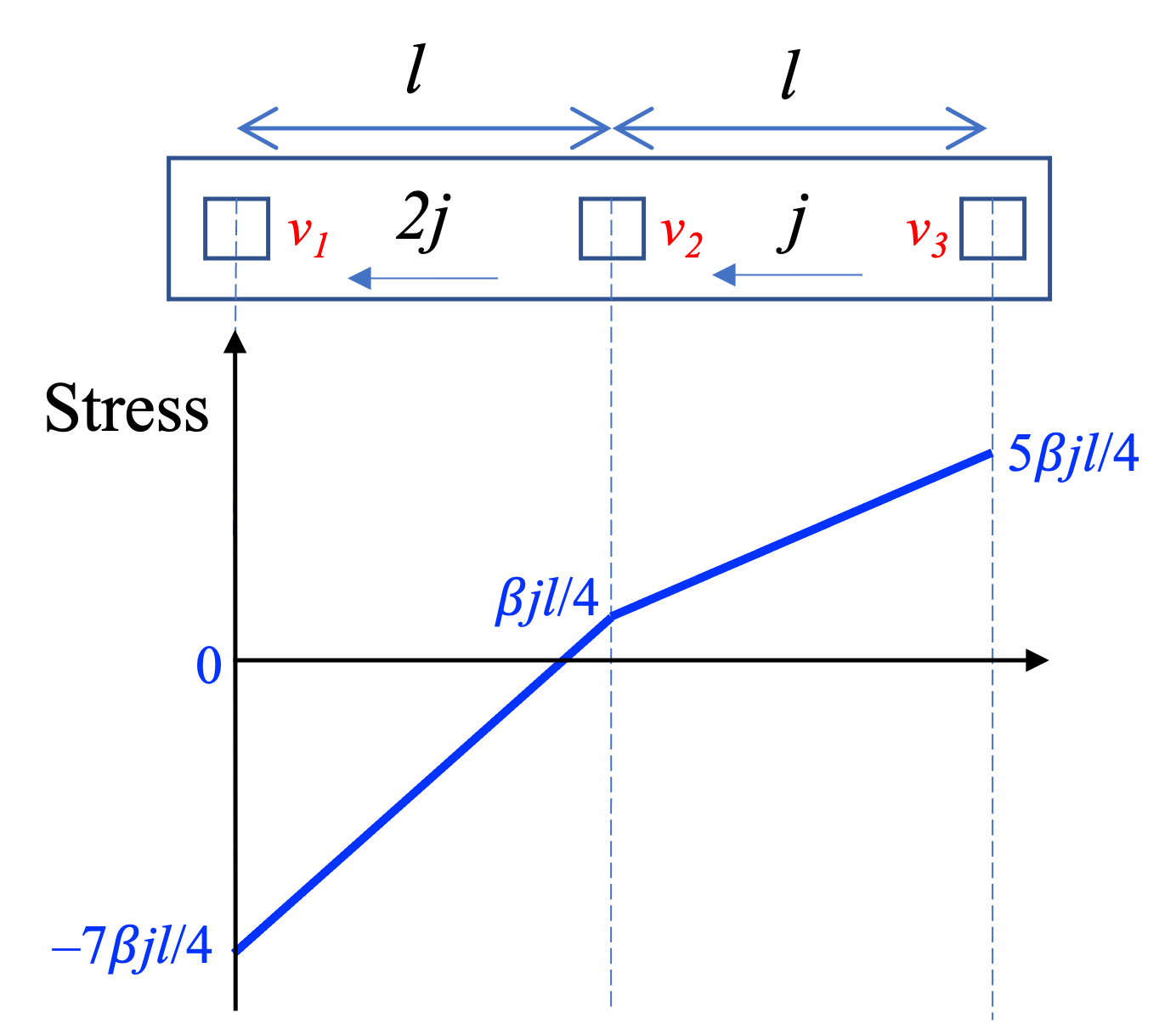}
\vspace{-3mm}
\caption{A two-segment interconnect, where the segment current densities are $j$ and $2j$; current directions correspond to electron current. Due to stress accumulation, the steady-state stress in the segment with \textit{lower} current density has a \textit{higher} positive tensile stress (which can nucleate a void).}
\Description{A two-segment interconnect, where the segment current densities are $j$ and $2j$; the current directions correspond to electron current. The steady-state stress in the wire shows a higher value in the segment with lower current density due to stress accumulation.}
\vspace{-4mm}
\label{fig:vianodevector}
\end{figure}

Conventional EM analysis is well known to be flawed for multisegment interconnects. As observed in~\cite{parkvianode:10}, measured data shows that when two segments conduct different currents in the same direction (as in Fig.~\ref{fig:vianodevector}), the segment that is closer to the cathode breaks down sooner even if it has a lower current density.  This phenomenon is easily explained using physics-based approaches that model stress build-up over multiple segments.

\subsection{EM equation formulation}
\label{sec:em_equations}

\noindent
\textbf{The EM PDE.}
A thorough treatment of physics-based EM analysis was provided in~\cite{kor:93} to analyze the stress that results from the three sources outlined in Section~\ref{sec:introduction}. This analysis can be used to analyze the nucleation and growth of voids in a wire. Although stress is a three-dimensional phenomenon, in interconnect wires where the length is the dominant dimension, a one-dimensional (1D) approximation suffices, assuming uniformity in the other two dimensions. A single interconnect segment of length $L$ injects electron current at a cathode at $x=0$ towards an anode at $x=L$.  The temporal evolution of EM-induced stress, $\sigma(x,t)$, at any point $x$ in the segment is modeled by the PDE:
\begin{align}
\frac{\partial \sigma}{\partial t} &= 
  \frac{\partial }{\partial x} \left [
         \kappa \left ( \frac{\partial \sigma}{\partial x} + G
                \right) \right ]
\label{eq:Korhonen's_eqn}
\end{align}
Here,  $G = \beta j$ and $\kappa = D_a {\mathcal B} \Omega/(kT)$, where $\beta = (\Zeff e \rho)/\Omega$, $j$ is the current density in   the wire, $\Zeff$ is the effective charge number, $e$ is the electron charge, which is negative, $\rho$ is the resistivity, $\Omega$ is the atomic volume for the metal,  ${\mathcal B}$ is the bulk modulus of the material, $k$ is Boltzmann's constant, $T$ is the temperature, and $D_a=D_0 e^{-E_a/kT}$ is the diffusion coefficient, with $E_a$ being the activation energy.

The sign convention for $j$ is in the direction of electron current, i.e., opposite to conventional current. Of the two terms on the right-hand side of~\eqref{eq:Korhonen's_eqn}, the second contains $G$ represents atomic flux attributable to the electron wind force; the first contains the stress gradient $\frac{\partial \sigma}{\partial x}$ that accounts for the flux related to the back-stress. The sum, $(\partial \sigma/\partial x +  G)$, is proportional to the net atomic flux. 

\begin{figure}[tb]
\centering
\includegraphics[width=\linewidth]{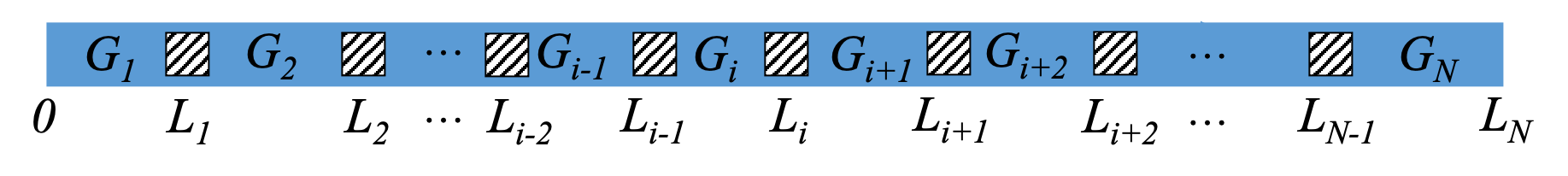}
\vspace{-6mm}
\caption{A multisegment interconnect line with different current densities in each segment and potentially nonuniform segment lengths.}
\Description{A multisegment interconnect line with different current densities in each segment and potentially nonuniform segment lengths.}
\vspace{-4mm}
\label{fig:multiseg}
\end{figure}

In an $N$-segment interconnect line of length $L_N$, shown in Fig.~\ref{fig:multiseg}, currents may be injected or drawn at intermediate points $x=L_i$ ($i=1,\cdots,N-1$) through vias, and each segment $i$ is associated with a current density $j_i$ and a corresponding parameter $G_i$. Stress evolution $\sigma_i(x,t)$ in each segment is described by the PDE~\eqref{eq:Korhonen's_eqn}, and the system of equations is supplemented by a set of boundary conditions (BCs) that relate the stress at various segments, as well as an initial condition $\sigma(x,0)$ for the line stress at time $t=0$. 

\noindent
\textbf{Initial conditions.}
The initial stress in a wire, prior to the application of current, is often taken to be zero in many research papers. However, strictly, the initial stress is the residual thermomechanical stress that remains in the wire due to differentials in the CTEs of the material of the wire and its surrounding materials.  In practice, this can be taken to be a constant value, $\sigma_T$, that can be analyzed by determining the residual stress. In principle, even under a reasonable 1D approximation, this stress may vary along the length a wire, but realistically, since voids tend to appear at the ends of line segments or at vias, initial values of $\sigma_T$ at the these points, if available, may be used to determine stress evolution until nucleation.

\noindent
\textbf{BCs: Nucleation phase.}
Since the terminal points of the line act as blocking boundaries, the BCs dictate that the atomic flux will be zero over all time at the end points $x=0$ and $x=L_N$, i.e.,
\begin{align}
\frac{\partial \sigma_1(0,t)}{\partial x} + G_1  = 0 \; ; \;
\frac{\partial \sigma_N(L_N,t)}{\partial x} + G_N  = 0
\label{eq:BC_multi_terminal}
\end{align}
At intermediate points $x=L_1,\cdots,L_{N-1}$, the BCs ensure the continuity of stress and the conservation of atomic flux~\cite{Sun18}, i.e. for $i=1,\cdots,N-1$,
\begin{align}
\sigma_i(L_i,t) &= \sigma_{i+1}(L_i,t)
\label{eq:BC_continuity_2}
\\
\frac{\partial \sigma_i(L_i,t)}{\partial x} + G_i &= 
\frac{\partial \sigma_{i+1}(L_i,t)}{\partial x} + G_{i+1}
\label{eq:BC_internal_2}
\end{align}


\noindent
\textbf{BCs: Postvoiding phase.}
In the postvoiding phase, the dynamic evolution of stress continues to follow the PDE~\eqref{eq:Korhonen's_eqn}, but the boundary condition at the location of the void nucleation will be different. Specifically, at the point of nucleation the stress will be zero, since a void, by definition, asserts no stress. Due to temporal continuity, around the void, the stress in the wire remains at its value just prior to void formation. Since stress must still be spatially continuous, it will exhibit a very large gradient from zero stress (in the void) to the pre-existing stress (around the void) over a very small distance, defined as the ``effective thickness of the void interface'' $\delta$, and being much smaller than the lengths of the interconnect segments~\cite{Sukharev16}. Then, if the location of void nucleation is $x_v$, the Robin-type boundary condition at $x_v$ is:
\begin{equation}
\frac{\partial\sigma(x_v,t)}{\partial x}=\frac{\sigma(x_v,t)}{\delta}
\label{eq:BC_pvd}
\end{equation}
Note that this boundary condition is also listed in~\cite{kor:93}, but instead of the void interface thickness $\delta$, the line width $w$ is used (which is also much smaller than the segment lengths in a 1D framework).

For regions (terminals or intermediate points) of the interconnect where a void is not nucleated, the boundary conditions remain the same as in the nucleation phase.

\subsection{Solving the EM equations}

\noindent
There are two modes in which the solution of the EM equations is useful: (i)~steady-state analysis, which determines the equilibrium stress in an interconnect when all forces are balanced, and (ii)~transient analysis, which determines the evolution of stress as a function of time until steady-state is achieved.

\subsubsection{Steady-state analysis}

\noindent
Steady-state analysis of the PDE involves zeroing out all derivatives with respect to time.  This results in a second-order differential equation in the distance variable $x$, which has a stress solution that is piecewise linear in $x$ (i.e., $d \sigma_i/dx = -\beta j_i$ along the length of wire $i$ carrying current $j_i$), and, by the boundary conditions, continuous at the vias.  Conventionally, it is assumed that the stress grows monotonically till it reaches the steady-state stress, which is considered to be the highest achievable stress in a wire:\footnote{However, this is not necessarily true for a multisegment line: see \textit{Caveat~4}.} if this value is below the critical stress, then the wire is considered immortal over all of its segments. In fact, an alternative view of the Blech criterion is that it arises from the steady-state solution of a single-segment wire: if the stress is then below $\sigma_c$, then the wire is immortal. For multisegment wires, steady-state analysis is thus a generalization of the Blech criterion~\cite{Shohel21a,Shohel21arxiv}.  

\begin{figure}
\centering
\includegraphics[width=0.7\linewidth]{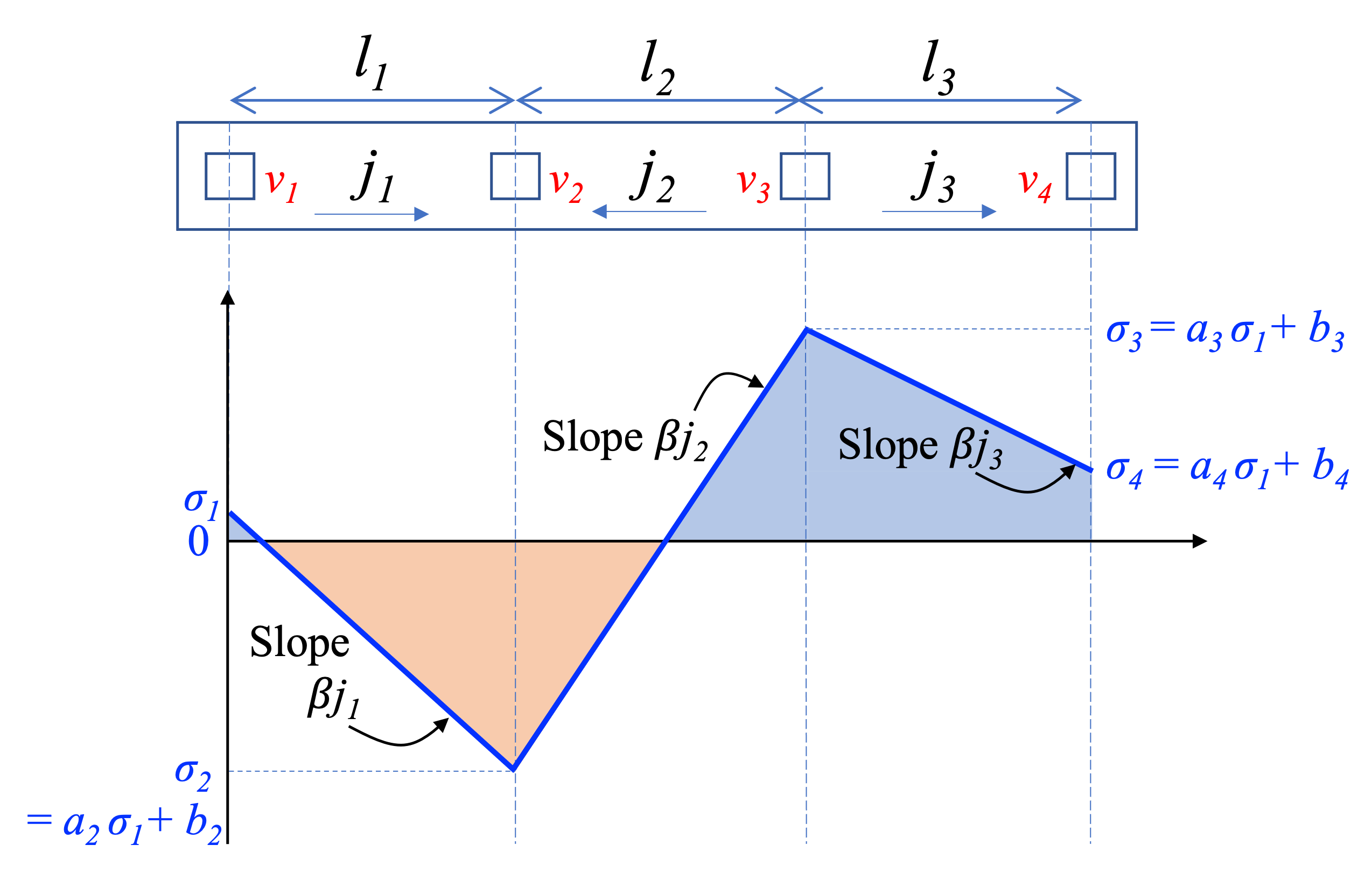}
\vspace{-3mm}
\caption{A three-segment example to illustrate the intuition behind fast
steady-state computation.}
\Description{A three-segment example to illustrate the intuition behind fast
steady-state computation.}
\vspace{-4mm}
\label{fig:3seg_example}
\end{figure}

Explanations of EM are often mired in the solutions of the partial differential equations presented in the previous section. We present an intuitive view of the solution of the steady-state problem. Fig.~\ref{fig:3seg_example} shows an example of a three-segment line and the steady-state stress, $\sigma_i$, for $i = 1, \cdots, 4$, at at each vertex $v_i$. As stated above, the slope of each segment is proportional to the current density in the segment; note that the sign of the slope in the middle segment is the opposite to that in the other segments since the direction of current flow is also the opposite to other segments.

Therefore, if we knew $\sigma_1$, the other stress values could be computed based on the known slopes: $\sigma_2$ is lower than $\sigma_1$ by $\beta j_1 l_1$; $\sigma_3$ is higher than $\sigma_2$ (due to the change in current direction) by $\beta j_2 l_2$; and $\sigma_4$ is reduced from $\sigma_3$ by $\beta j_3 l_3$.  In other words, given all $j_i$ values, all stresses may be written in terms of $\sigma_1$ as shown in the figure (as we will show later, this is equally true for tree structures and meshes), where the values of $a_2, a_3, a_4, b_2, b_3, b_4$ in the figure depend on the currents in the lines. However, $\sigma_1$ is as yet unknown.

To find $\sigma_1$, we simply appeal to the concept of conservation of mass:  although atoms are transported along the wire, there is no change in the total number of atoms in the wire due to EM. For wires of equal width and thickness, as in this example, the integral of stress over the region is zero~\cite{Haznedar06}. Since all $\sigma_i$ values are expressed in terms of $\sigma_1$, this integral depends purely on $\sigma_1$: equating it to zero provides $\sigma_1$, and therefore, $\sigma_2, \cdots, \sigma_4$.

Various techniques for steady-state analysis have been presented in the literature~\cite{Demircan14,Gibson14,Najm20,Shohel21a,Shohel21arxiv} that can solve the steady-state problem in linear time. The key intuition behind each of these is that under certain assumptions, the stress difference between two nodes is proportional to the difference in voltages between the nodes~\cite{Haznedar06}. To see this, consider a wire segment $i$ whose length, width, and thickness are $l_i$, $w_i$, and $h_i$, respectively, between circuit nodes $a$ and $b$.  Then its resistance $R_i = \rho l_i/(w_i h_i)$, where $\rho$ is the wire resistivity. By Ohm's law, the electron current $j_i$ from node $b$ to node $a$ is given by:
\begin{align}
j_i = \frac{V_b - V_a}{R_i (w_i h_i)} = \frac{V_b - V_a}{\rho l_i}
\end{align}
Since the stress on segment $i$ varies linearly with a slope of $-\beta j_i$,
\begin{align}
\sigma^b - \sigma^a &= -\beta j_i l_i = -\frac{\beta}{\rho} (V_b - V_a)
\label{eq:linearstress}
\end{align}
The above result has three important consequences:
\begin{enumerate}[itemsep=0pt,leftmargin=*,topsep=0pt]
\item The computation of steady-state stress follows immediately from DC voltage analysis: under a set of current excitations, if the voltage at each node is computed, then the values of each voltage, relative to a datum node, are known from the system of difference equations~\eqref{eq:linearstress}.  To compute the value at the datum, we appeal to the conservation of mass equation, as in Fig.~\ref{fig:3seg_example}, and this can be shown to be a linear sum of all node voltages~\cite{Demircan14,Najm20,Sun16,Shohel21arxiv}.
\item According to Kirchhoff's voltage law (KVL), the sum of voltages around any cycle is zero; Eq.~\eqref{eq:linearstress} implies that the stress around any cycle must also be zero.  As a result, even with interconnect systems that have cycles, it is sufficient to create a tree over the interconnect graph, and computing the stresses based on voltage values in the tree.
\item A small change to Eq.~\eqref{eq:linearstress} shows that the $j_i l_i$ product is proportional to $I_i R_i$, where $I_i = (V_b-V_a)/R_i$ is the current through the wire. Thus, the stress difference between two points is proportional to the voltage drop in the wire.  In other words, optimizing a power grid to control the IR drop within the grid is consistent with reducing stress and increasing EM robustness.  While this is intuitively consistent with the idea that both IR drop reduction and EM robustness require wire widening, the quantitative relationship above allows for steady-state EM optimization to be natively incorporated into DC optimization of a power grid.
\end{enumerate}

\begin{figure}[t]
\centering
\includegraphics[width=\linewidth]{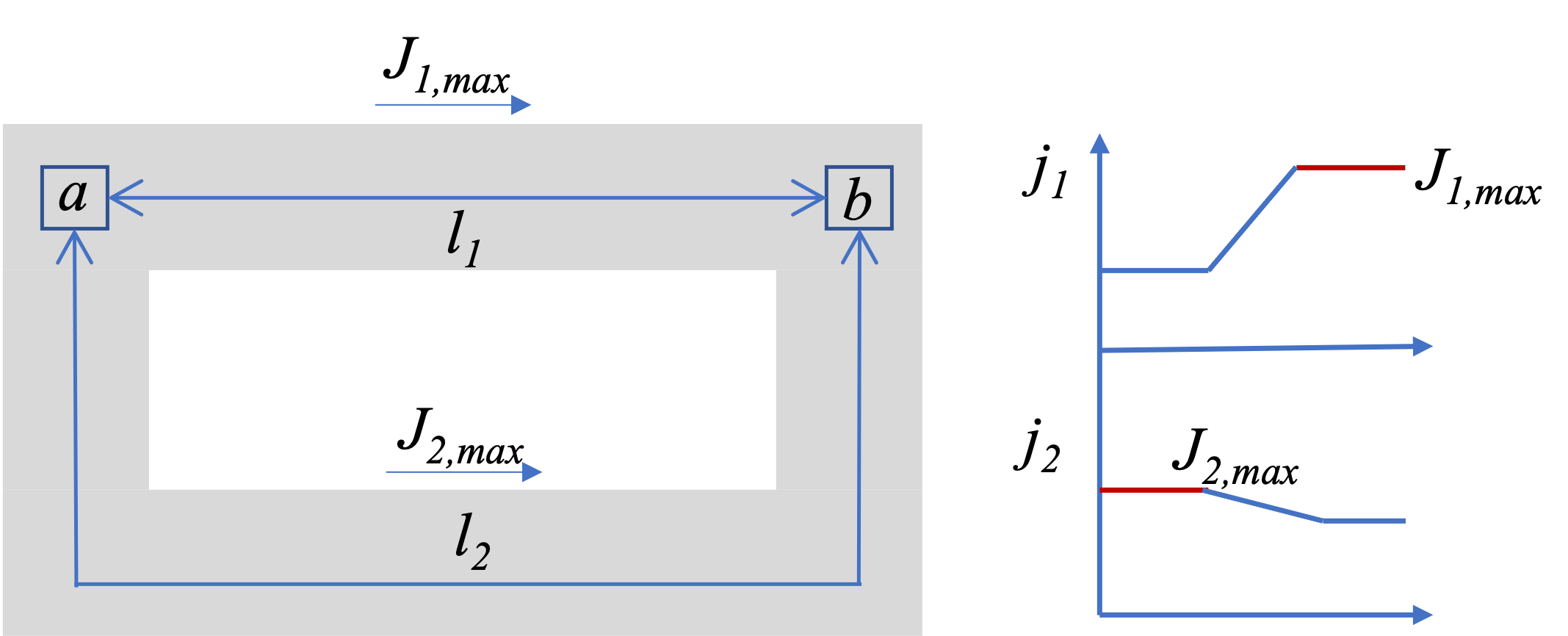}
\vspace{-3mm}
\caption{A simple example to illustrate that the KVL relationship may not hold across multiple time points.}
\Description{A simple example to illustrate that the KVL relationship may not hold across multiple time points.}
\label{fig:DCcaveat}
\vspace{-4mm}
\end{figure}

\noindent
\textit{\underline{Caveat~2}}
Item (2) above is based on KVL, which is true when the voltages and currents all correspond to values at a specific snapshot in time.  In such a case, $\sum_{\text{edges $i$ in a cycle}} I_i R_i = 0$. However, for EM analysis, the current value in a wire segment may be taken to be the \textit{worst-case value over a period}: if so, the KVL relationship does not hold true, and the stress around a cycle is not zero.  For the two-segment cycle involving nodes $a$ and $b$ in Fig.~\ref{fig:DCcaveat}, $J_{1,max}$ and $J_{2,max}$ are chosen as the largest values over a time period: since these occur at different times in the waveform, they may not obey KVL.
Moreover, in this case, conservation of mass does not hold across different time points.  Therefore, the solution of the steady-state EM problem in this realistic scenario, where the peak currents occur at different time points, do not obey KVL and therefore cannot utilize \eqref{eq:linearstress} to remove cycles from a graph: instead, a worst case must be found incorporating the longest path in the natural graph that arises from these difference constraints.

\noindent
\textit{\underline{Caveat~3}}
Several prior works have stated that the maximum stress in a wire is proportional to the largest sum of the $j_i L_i$ products on any path between two nodes in an interconnect tree~\cite{vema,ala:05}. This is not true:  the normalization associated with the mass conservation constraint, illustrated in Fig.~\ref{fig:3seg_example}, creates a denominator with the $j_k L_k$ sum over all segments in the tree. Hence, the stress does not vary linearly with the sum of these $j_i L_i$ products on the path.

\subsubsection{Transient analysis}
\label{sec:transient_analysis}

\noindent
The steady-state formulation is useful for predicting the immortality of an interconnect. Quite simply, if the steady-state stress lies below the critical stress, the wire is immortal; if not, it is potentially mortal.  A method for performing this analysis in an organized manner, using multiple filtering steps to efficiently weed out EM-safe nets, is presented in~\cite{Mishra16}. For a potentially mortal wire, it is necessary to perform transient analysis to analyze the evolution of stress over time to verify whether the stress exceeds the critical stress over the lifetime of the chip.  For a single-segment interconnect, several approaches have been proposed for predicting the evolution of stress over time, including an infinite series solution in~\cite{kor:93}.  For multisegment interconnect structures, several classes of methods have been proposed:

\begin{itemize}[itemsep=0pt,leftmargin=*,topsep=0pt]
\item \textit{Numerical time-domain solutions} may use finite element or finite difference solvers using commercial tools such as COMSOL or ABAQUS, or customized efficient solutions~\cite{Chatterjee18,Axelou22,Stoikos23} that solve the linear time-invariant (LTI) system in the time domain

\item \textit{Frequency-domain methods} formulate the problem using a state-space representation of the LTI system and reduce the system matrices in the frequency domain using Krylov subspace methods~\cite{Cook18}.

\item \textit{Infinite series methods} develop closed-form solutions to the PDE that meet the boundary conditions. In \cite{hua:14,Chen2016}, the infinite series solution is truncated it to a suitable number of terms. However, (a) it is very difficult to calculate and write down the infinite series for more than 3-4 segments; (b) it is practically impossible to obtain expressions parameterized by a general number of segments; (c) it is difficult to know how many terms the series should be truncated to while maintaining sufficient accuracy. 

\item \textit{Stress-wave (or reflection-based) methods}~\cite{Shohel21b,Shohel25} achieve low computation times through a parallel with a stress wave emanating from each end-point or via in a multisegment interconnect structure and is reflected at the boundary points. In~\cite{Shohel21b,Shohel25}, this approach is only applicable to straight multisegment lines without branches, but extensions to consider branches are possible.  The idea can be shown to be equivalent to an infinite series method, but overcomes the three disadvantages listed above by building the series from the bottom up and providing a theoretically sound criterion for terminating the series, whose terms correspond to reflections of each fundamental wave, providing a \textit{linear-time solution} to the transient analysis problem.

\item \textit{Equivalent RC circuit methods}~\cite{Najm21,Shohel23} draw an analogy between the stress PDE and the heat equation and, similar to the thermal-electrical equivalence, propose a stress-electrical equivalence, representing the solution of the spatially discretized PDE as the analysis of an RC circuit. In~\cite{Shohel23}, for interconnect tree structures, this RC circuit is reduced in \textit{linear time} in the frequency domain using model order reduction to yield a small reduced-order model that can be used to provide a very efficient solution.
\end{itemize}
The stress-electrical analogy opens the doors to exploiting the considerable machinery available in the field of electrical network analysis~\cite{Celik02,Sapatnekar04}. Discretizing the interconnect wires into elements of size $\Delta x$, the electrical circuit is built, based on the following equivalences:
\begin{itemize}[itemsep=0pt,leftmargin=*,topsep=0pt]
    \item The voltage $V_i$ at the node $i$ in the circuit is equivalent to the circuit stress $\sigma_i$ at the corresponding node.
    \item A node $i$ in the electrical equivalent circuit has a grounded \textit{stress capacitance} of $(\Delta x \cdot w \cdot h)$, where $w \cdot h$ is the wire cross-section.
    \item Nodes $j$ and $k$, corresponding to adjacent elements, are connected by a \textit{stress resistance} of value $\frac{1}{\kappa} \frac{\Delta x}{(w \cdot h)}$, where $\kappa$ is defined in~\eqref{eq:Korhonen's_eqn}.
    \item At each end point or via node $i$, a \textit{current source} to ground is inserted to obey Kirchhoff's current law (KCL).  In other words, the current to ground through the source is the algebraic sum of all interconnect currents entering node $i$.
\end{itemize}
We can observe that the {\em stress resistance} has the familiar form of the electrical resistance of a conducting material of length $\Delta x$, cross-section $(w \cdot h)$ and resistivity $\frac{1}{\kappa}$, while the {\em stress capacitance} is equal to the volume $(\Delta x  \cdot w \cdot h)$ of the discretized element.

\begin{figure}[tb]
\centering
(a) \subfloat{\includegraphics[width=0.9\linewidth]{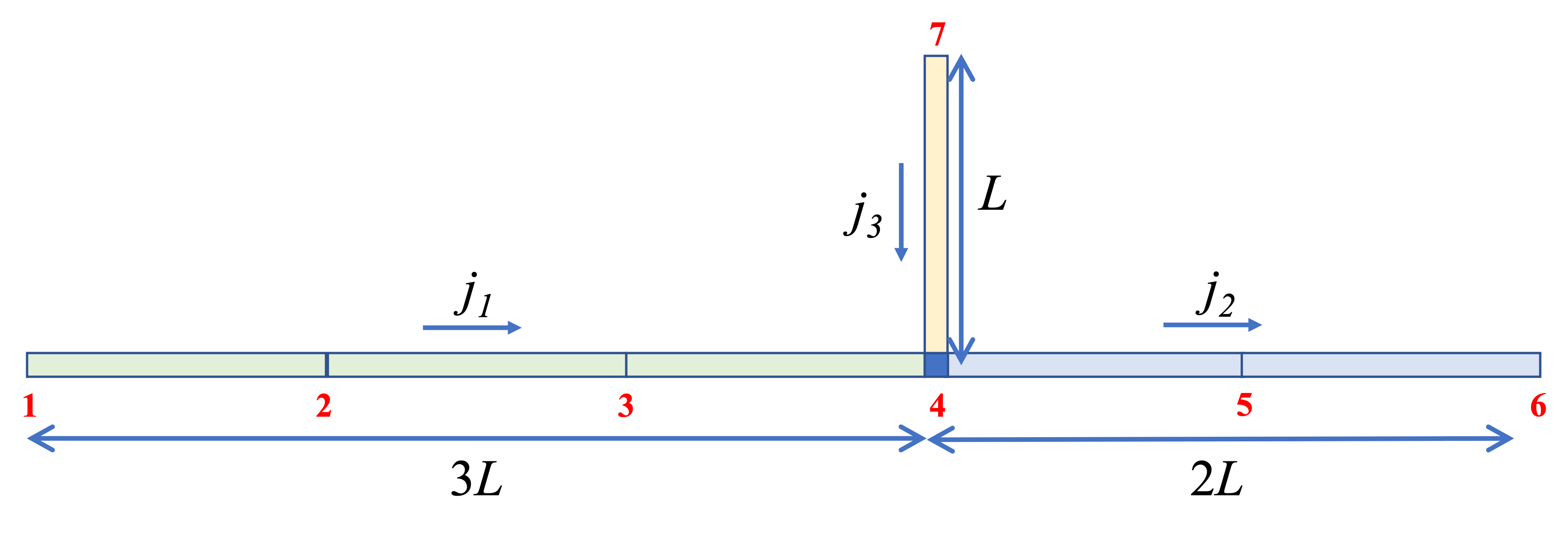}}

\vspace{-6mm}
(b) \subfloat{\includegraphics[width=0.9\linewidth]{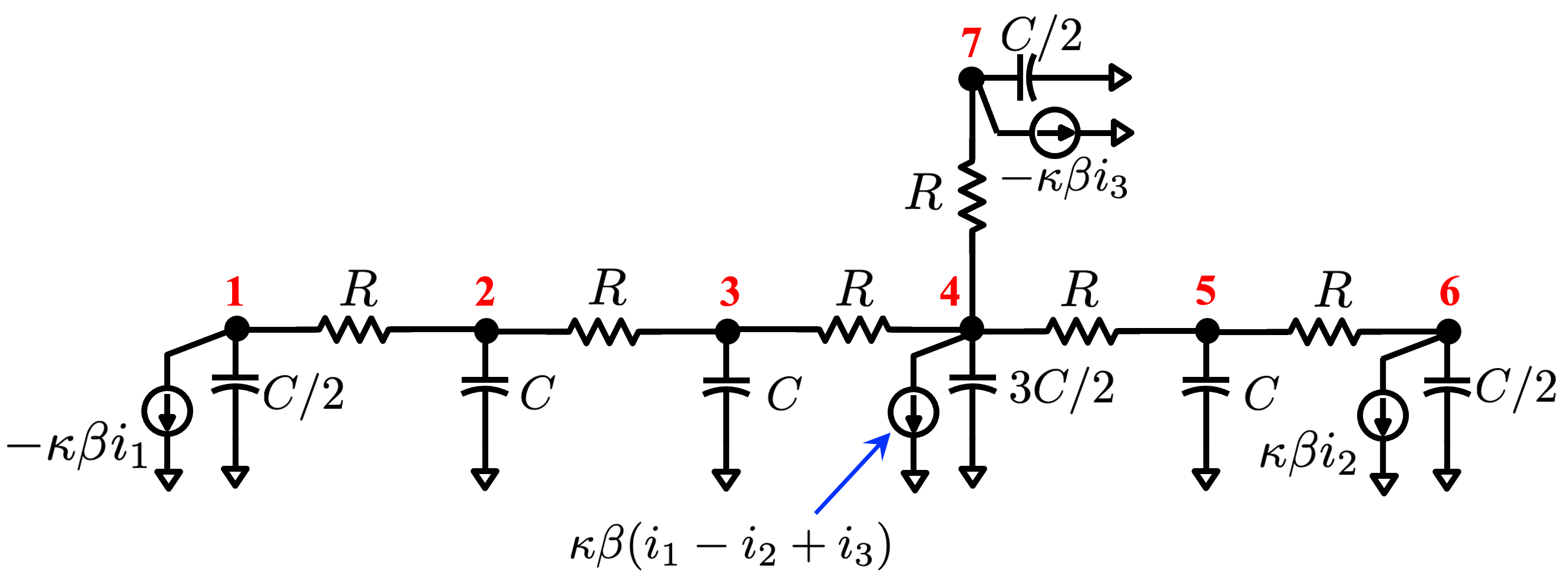}}
\vspace{-3mm}
\caption{(a)~A three-segment tree discretized into elements. 
(b)~The equivalent RC circuit with excitations.}
\Description{(a)~A three-segment tree discretized into elements. 
(b)~The equivalent RC circuit with excitations.}
\vspace{-4mm}
\label{fig:example_line_3seg}
\end{figure}

We illustrate a mapping of the stress problem to an equivalent electrical network for the prenucleation phase, based on the above principles.  Our illustrative example is the three-segment tree of Fig.~\ref{fig:example_line_3seg}(a), with uneven segment lengths, equal segment widths and thicknesses, and with current densities $j_1$, $j_2$, and $j_3$.

After discretization into elements of length $L$, from the stress-electrical equivalence, 
we can build the equivalent RC circuit shown in Fig.~\ref{fig:example_line_3seg}(b), where each discretized element has identical resistance $R$ and identical capacitance $C$.  Current sources are added to nodes that correspond to segment end-points: the source value for each segment end node is equal to the algebraic sum of the stressing currents on segments incident on the node. 

\noindent
\textit{\underline{Caveat~4}}
It is commonly, but incorrectly, assumed that the transient stress must grow monotonically over time, so that the maximum stress magnitude in a wire corresponds to the steady-state solution.  This is provably true for a single-segment interconnect, but may not be true in a multi-segment wire, where stress accumulates over time over multiple interconnect segments.  In such a case, it is possible to build scenarios where the impact of accumulated stress from other wire segments is small and the stress build up in a single segment is much larger.  Over time, as the other wire segments reach steady state, this may bring in a stress component with an opposing sign that brings down the initial stress build-up. Therefore, it is possible to nucleate a void during the transient as the stress exceeds $\sigma_c$, even though the steady-state stress lies below $\sigma_c$.

\section{Circuit-level EM failures}
\label{sec:circuits}

\noindent
The solutions covered so far discuss the computation of stress in an individual interconnect, and determine when an individual interconnect might fail.   At the circuit level, it is necessary to develop criteria for failure when a single wire or net fails.

Early approaches~\cite{Kitchin96,Li11} use the ``weakest link'' criterion to evaluate system failure: for a system with $K$ elements, the system is assumed to be functional if \textit{all} $K$ elements are EM-failure-free.  If $F_i(t)$ is the failure probability of element $i$ at time $t$, then the probability $F_{chip}(t)$ of chip failure is then given by:
\begin{align}
F_{chip}(t) = 1 - \prod_{i=1}^K (1 - F_i(t))
\end{align}
The weakest link approach assumes that every system element must be functional in order for the system to work correctly. In practice, many systems have a significant amount of redundancy, so that a failure in one wire does not cause system failure. For example:

\begin{figure}[b]
\centering
\vspace{-10mm}
\includegraphics[height=5cm]{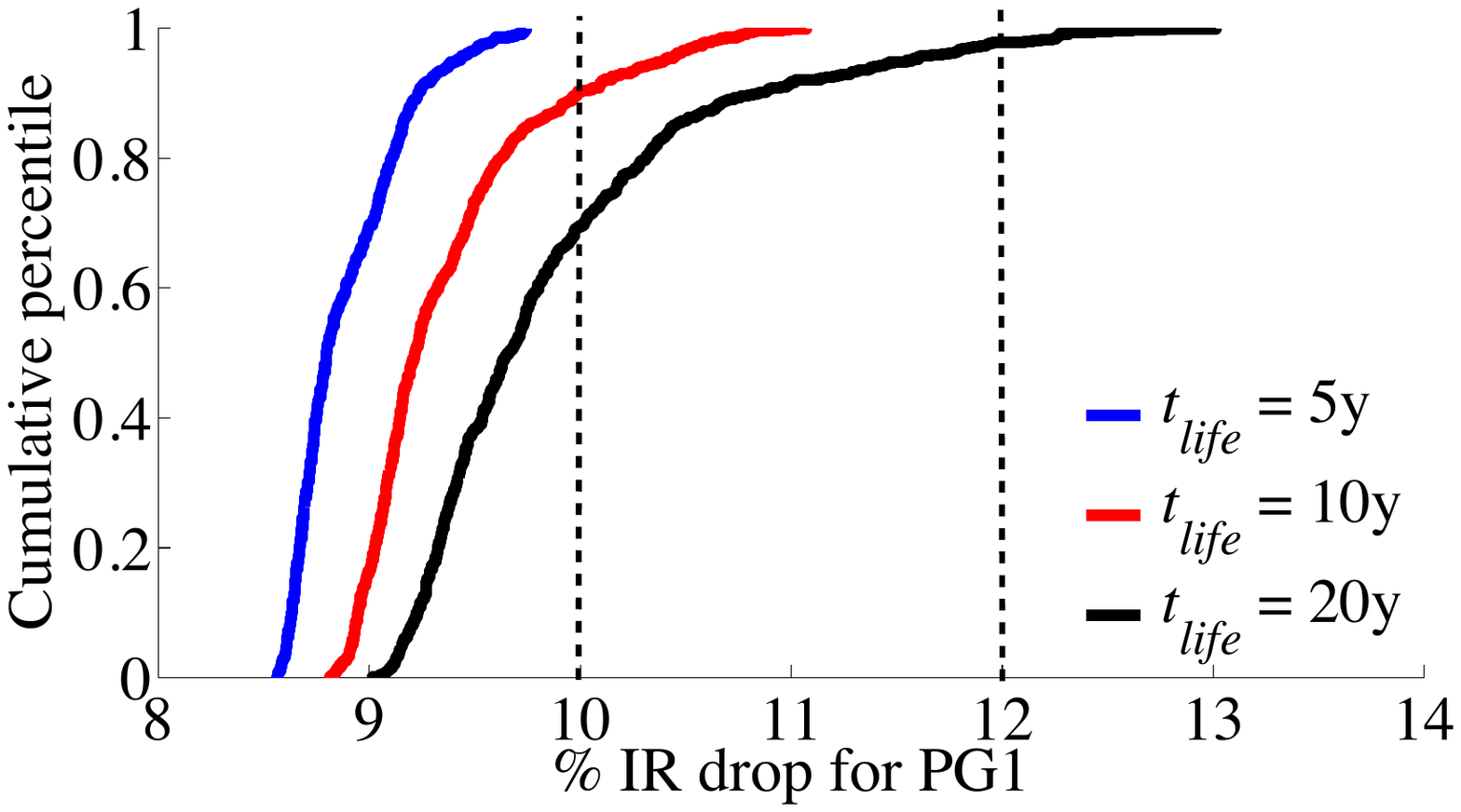}
\vspace{-10mm}
\caption{IR drop CDF for PG1 for various lifetimes, $\tlife$~\cite{vivek:dac}.}
\Description{IR drop CDF for PG1 for various lifetimes, $\tlife$~\cite{vivek:dac}.}
\label{fig:cdfpg5}
\end{figure}

\begin{itemize}[itemsep=0pt,leftmargin=*,topsep=0pt]
\item 
In power grids~\cite{vivek:dac}, the failure of a single link may increase voltage drops, but the system may still function correctly.  Fig.~\ref{fig:cdfpg5}~\cite{vivek:dac} shows the CDF of the maximum IR drop in the IBM benchmark pg1, for various values of the lifetime, $\tlife$,  corresponding to the typical product lifetime values for mobile (5 years), computing (10 years), and automotive (20 years) applications, respectively. It can be seen that as time progresses, the probability of the IR drop exceeding a specified threshold (e.g., 10\% or 12\%) increases. A similar observation was subsequently made in~\cite{najm2:13}, and experimentally demonstrated in~\cite{Zhou20}. Power grids also use via arrays that are a point of EM susceptibility, and the redundancy of these arrays leads to system resilience to EM~\cite{Mishra17}.

\item Clock meshes have multiple clock drivers connected to a meshed high-level clock distribution network.  In~\cite{JainIRPS15}, it was shown that such a clock mesh can withstand multiple EM failures due to redundancy in the network.
\end{itemize}
In such redundant systems, when one interconnect segment fails, the current load through other wires increases, so that the failure probabilities of the other segments are increased after each failure.


%% file: sec/4-Open-Analysis.tex
\vspace{-4mm}
\section{Some open problems in EM analysis}
\label{sec:open-analysis}

\noindent
We discuss two open problems in EM analysis: (a)~evaluating the impact of variations, and (b)~calibrating EM test structures to characterize parameters required to enable physics-based analysis.

\subsection{Impact of variations on EM}

\noindent
Performance metrics for integrated circuits are particularly susceptible to process, voltage, and temperature (PVT) variations, and the EM lifetime is no exception.  The Arrhenius term in~\eqref{eq:black} shows the exponential temperature dependence of MTTF that has been observed empirically; a similar term appears in the expression for the diffusion coefficient, $D_a$, that appears in the EM PDE~\eqref{eq:Korhonen's_eqn}.  The exponential dependence on process parameters is also seen in the same terms of these equations, and is primarily embedded in variations in the activation energy, $E_a$. This parameter is empirically seen to show a normal distribution, leading to a lognormal distribution for the MTTF.  Voltage variations are typically less serious and lead to linear changes in the stressing currents.

\textbf{Temperature variations} in an interconnect arise due to two causes: (a)~temperature rises across the chip due to the heat generated within transistors in nearby functional blocks, and (b)~Joule heating in a wire as it conducts electricity.  Both of these can lead to faster EM degradation~\cite{Kteyan22}.  Unfortunately, the first factor depends on the operating conditions of the chip: using worst-case temperatures leads to very pessimistic results in cases where these large temperatures occur infrequently, and finding a precise worst-case operating conditions is a difficult problem.  Joule heating can be modeled more easily and incorporated into EM models. In principle, thermomigration can also be a factor, but requires very significant thermal gradients that are not common across the span of a wire.

Another source of variations is the set of \textbf{current excitations} on chips in the field. Finding the true worst-case excitation for EM has been an open problem for many years.

\noindent
\textbf{Incorporating process variations into EM analysis.}
In the literature, the predominant approach to handling process variation effects is through Monte Carlo analysis. In~\cite{Chatterjee18}, such an approach is used around a fast finite-difference solver, adding further heuristics such as a TTF predictor for better computational efficiency.

Greater efficiencies are possible, leveraging the RC circuit interpretation (illustrated in Fig.~\ref{fig:example_line_3seg}) and the fast linear-time solutions for nominal stress under that model.  It can be observed that a change in the activation energy translates to a shift in the values of the element resistance $R$ and the current source magnitudes. Both quantities depend on $\kappa$, and hence vary exponentially with $E_a$. In the frequency domain, the system equations are~\cite{Shohel23}:
\begin{align}
(G + s C) \mathbf{V}(s) = \beta \kappa \mathbf{J} u(s)
\label{eq:orig}
\end{align}
where $G$ and $C$ are the conductance and capacitance matrices, $\mathbf{J}$ is constructed from the current sources, $\mathbf{V}$(s) is the response (stress), and $u(s)=1/s$ corresponds to the unit step function.  Let $\Delta \mathbf{V}(s)$ be the shift in $\mathbf{V}(s)$ as the activation energy goes from $E_a$ to $E_a + \Delta E_a$. This change shifts $\kappa$ to $\kappa + \Delta \kappa$, changing each conductance, and hence perturbing $G$ to $G + \Delta G$. The new circuit equations are:
\begin{align}
(G + \Delta G + s C) (\mathbf{V}(s) + \Delta \mathbf{V}(s)) = \beta (\kappa + \Delta \kappa) \mathbf{J} u(s)
\label{eq:perturbed}
\end{align}
Subtracting~\eqref{eq:orig} from \eqref{eq:perturbed}, and neglecting the product $\Delta G \Delta \mathbf{V}(s)$,
\begin{align}
(G + sC) \Delta \mathbf{V}(s) = \beta \Delta \kappa \mathbf{J} u(s) - \Delta G \mathbf{V}(s)
\end{align}
Under a normal distribution of $E_a$~\cite{Nair19} with variance var($E_a$), $\kappa$ is distributed lognormally with a mean of $\lambda$ times its nominal value, where $\lambda = (e^{\mbox{\scriptsize var}(E_a)/2(kT)^2} - 1)$. Since conductances are proportional to $\kappa$, the mean wire conductance $g_i$ is also $\lambda$ times its nominal value.

Therefore, taking the expectation of each side and using~\eqref{eq:orig}:
\begin{align}
(\lambda G + sC) E[\Delta \mathbf{V}(s)] = \lambda (\beta \kappa \mathbf{J} u(s) - G \mathbf{V}(s)) = \lambda sC \mathbf{V}(s).
\label{eq:expperturbed}
\end{align}

\begin{figure}
\centering
\includegraphics[width=0.77\linewidth]{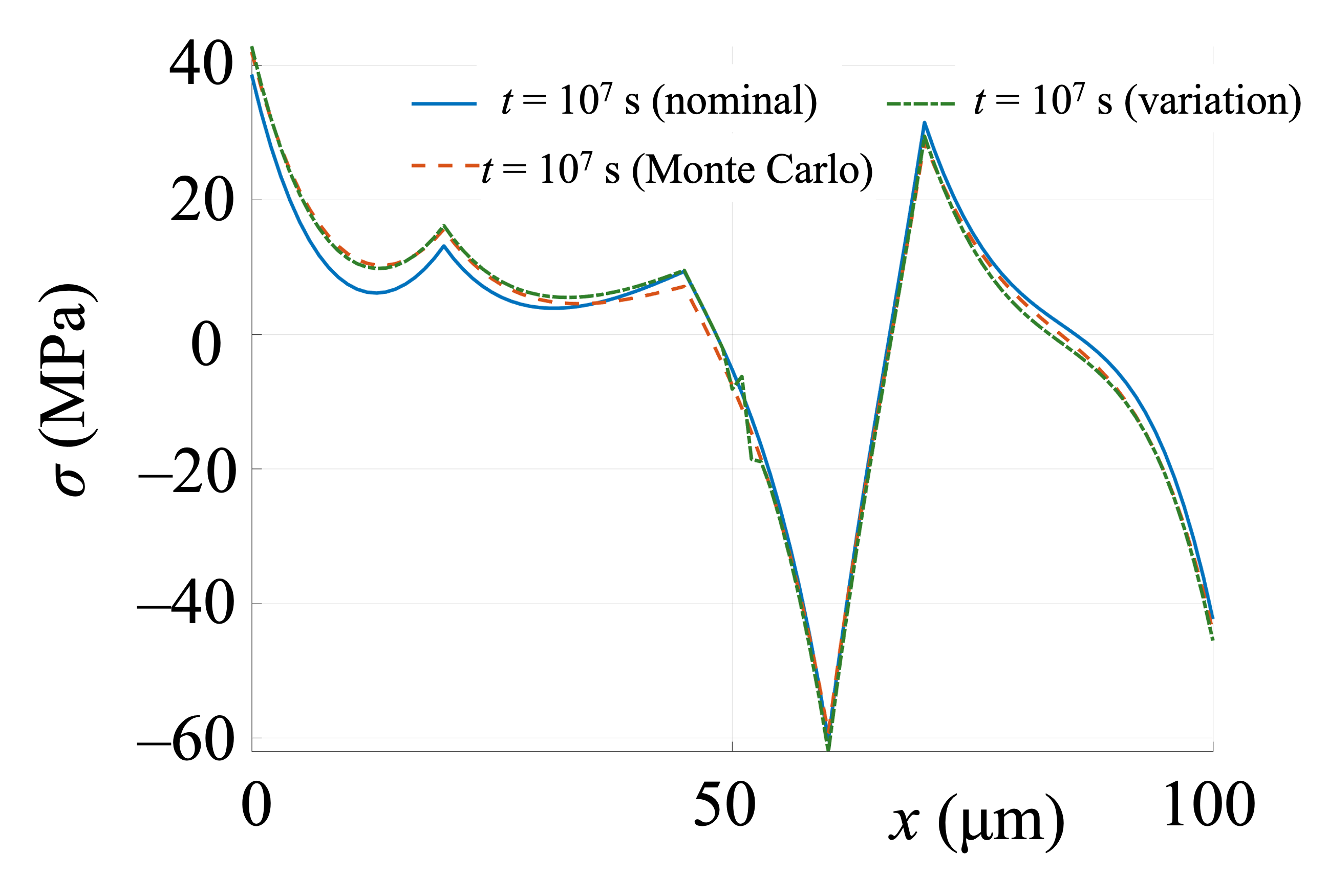}
\vspace{-2mm}
\caption{Fast variational analysis on a five-segment example.}
\Description{Results on fast variational analysis on a five-segment example.}
\vspace{-4mm}
\label{fig:variation}
\end{figure}

Using the Taylor series $\mathbf{V}(s) = \sum_{k=0}^\infty \mathbf{M}_k s^k$ and $\Delta \mathbf{V}(s) = \sum_{k=0}^\infty \mathbf{m}_k s^k$, and then matching the coefficients of $s^k$ on both sides of~\eqref{eq:expperturbed},
\begin{align}
G \mathbf{m}_0 = 0; \quad G \mathbf{m}_k = \lambda C \mathbf{M}_{k-1} - C \mathbf{m}_{k-1} \forall i>0
\end{align}
Further algebraic manipulations reveal that the moments of $\Delta \mathbf{V}(s)$ have the following simple relationship to those of $\mathbf{V}(s)$:
\begin{align}
\mathbf{m}_k = -k  \lambda \mathbf{M}_k
\end{align}
Thus, the moments of the mean stress under variation are obtained directly from the nominal moments. The application of this method is shown for a five-segment example in Fig.~\ref{fig:variation}, and shows excellent agreement between this prediction, built on top of~\cite{Shohel23}, and an expensive Monte Carlo solution.  The computation is very fast, almost identical to~\cite{Shohel23}, since its only additional cost is the multiplication of the $k^{\rm th}$ moment by $-k$. This enables linear-time computation of the mean stress and MTTF under variations, similar to~\cite{Shohel23}.

\subsection{Parameter calibration for physics-based EM}

\noindent
The contemporary approach to characterizing EM is based on measurements on a battery of test structures, corresponding to one-segment interconnects of constant length~\cite{lee:recovery}. These interconnects are subjected  to stress at high temperatures and high currents for rapid aging, and the distribution of lifetimes is reported. This distribution is typically lognormal: its mean (MTTF) and variance are extracted.  Innovative techniques use local heaters to raise the temperature of the metallization to over 300$^\circ$C, thus avoiding global heating of the chip that may also age the CMOS drivers, which could result in unpredictable currents flowing through the wires~\cite{zhou17}.

To translate EM lifetimes from stressed to normal operating conditions, we begin with the fundamental relation between the
EM lifetime and the lognormal variable. From \eqref{eq:black} and \eqref{eq:FF2z},
\begin{equation}
\sigma z = \ln(t_f) - \ln(A) + n \ln(j) - \frac{E_a}{k T}
\label{eq:acc_aging}
\end{equation}
Considering two stress conditions, an accelerated condition $a$ and a normal condition $b$, each is described by the same Black's equation parameters $A$ and $n$. For the same failure fraction, $z_a = z_b$, we can obtain the failure time under condition $b$ as~\cite{Jain16}:
\begin{equation}
t_{f,b} = \exp \left [ \ln (t_{f,a}) + \ln \left ( \frac{j_a^n}{j_b^n} \right )
    - \frac{E_a}{k} \left ( \frac{1}{T_a} - \frac{1}{T_b} \right ) \right ]
\end{equation}
where the subscripts $a$ and $b$ correspond to the accelerated and normal condition, respectively.

Most measurement efforts to date have focused on Black's law characterization, and there have been few efforts to calibrate physics-based models through measurements. A recent effort~\cite{Rothe25} is based on Korhonen's derivation of the infinite series solution for stress for a single-segment interconnect.  From~\cite{kor:93}, it can be shown that 
\begin{align}
	jL=\frac{\sigma_\textup{crit}}{\beta}  \Biggl( 0.5 - 4 \sum_{m=0}^{\infty} \frac{\textup{exp} \left( -(2m\pi + \pi)^2 \kappa \frac{t_\textup{life}}{L^2}\right)}{(2m\pi + \pi)^2} \Biggr)^{-1}
    \label{eq:jL-sigmacrit}
\end{align}
Based on measured data that plots $t_\textup{life}/L^2$ against $jL$, key parameters of the above equation can be derived. Specifically, based on these measured data points, it is possible to fit values of the parameters $\sigma_\textup{crit}/\beta$ and $\kappa$, and this provides a way of characterizing a single segment. To extend this approach to multisegment interconnects,~\cite{Rothe25} proposes to use these characterized values to determine the elements of the RC stress network for EM, as described in Section~\ref{sec:transient_analysis}. Expanding this research direction to show measured data on characterized test structures is an area of great value.

%% file: sec/5-Optimization.tex
\section{Some open problems in EM optimization}
\label{sec:open-optimization}

\noindent
One of the most effective ways to reduce EM-induced aging is to lower current densities. The difference between physics-based models and empirical models is that the former can capture the impact of accumulated stress over multiple wire segments, leading to better predictions of EM-susceptible regions in a chip.

\subsection{Incorporating EM into PDN optimization}

\noindent
EM overwhelmingly affects wires that carry unidirectional currents, such as analog biasing networks and power delivery networks (PDNs) that distribute the supply and ground signals over a chip.  Methods for optimizing the design of PDNs~\cite{Su03} include wire widening and topology selection, with the intent of reducing wire resistances and thereby reducing IR drop.  As described in Section~\ref{sec:physics} in the discussion of Eq.~\eqref{eq:linearstress}, there is a direct relationship between $I_i R_i$ and $j_i L_i$ for a wire $i$, raising the possibility of incorporating EM constraints related to $j_i L_i$ directly into a power grid optimization problem.  Accounting for the normalization of the denominator by a sum of $jL$ terms, constraining the steady-stage stress to lie below a specification results in the constraint, i.e.,
\begin{equation}
\sigma_i = \frac{\sum_{i \in \text{path}} j_i L_i}{\sum_{k \in \text{interconnect}} j_k L_k} \leq \sigma_c
\end{equation}
For a fixed interconnect topology with constant $L_i$ values, this leads to a linear constraint in the $j_i$ variables, easily incorporated in a PDN optimizer.  

As stated earlier, PDNs show redundancy and are unlikely to fail on the first EM failure: modeling and optimizing this, beyond initial efforts in~\cite{vivek:dac,najm2:13}, is also a promising research direction.

\subsection{Using reservoirs}

\noindent
A reservoir is a metal stub that lies upstream of the cathode and carries no current.  On EM-susceptible wires, the presence of a reservoir provides the availability of additional metal atoms: the effectiveness of reservoirs has been demonstrated both theoretically and experimentally~\cite{Dion01,Oates17}.  Within the reservoir, since there is no current, the only forcing function is the ``back-stress'' due to the concentration gradient between the stub and the cathode: since the reservoir is upstream of the cathode, this is actually a \textit{forward-stress} that supplies atoms to the depleted regions. Optimizing a layout to incorporate reservoirs, while accounting for the routing overhead of the stub, is an open physical design problem.

\subsection{Leveraging EM recovery}

\noindent
Realistically, few electronic systems are constantly turned on, and a typical experience is for the system to be powered off or placed in sleep mode from time to time. When a wire no longer carries current, there is no electron wind, and the back-stress leads to the reverse migration of metal atoms, leading to recovery. To date, few efforts have examined this effect~\cite{Huang16} or used it for optimization.  In the context of transistor reliability mechanisms, such recovery mechanisms have been well recognized, and methods for leveraging recovery have been proposed, e.g., using the idea of circadian rhythms in~\cite{Gupta12,Gupta13}: similar approaches could also be used for EM.

%% file: sec/6-Conclusion.tex
\section{Conclusion}
\label{sec:conclusion}

Research in electromigration is at an exciting juncture today. EM problems are of great importance to contemporary systems and emerging HI technologies. Conventional semi-empirical methods are known to have significant errors, and physics-based methods have reached a high level of maturity, enabling linear-time solutions to steady-state and transient analysis problems.  To move to the next step, it is necessary to connect EM characterization methods to the needs of physics-based methods and to solve several open problems in EM analysis and optimization.